\documentclass{rstransa}%%%%where rstrans is the template name

\titlehead{Research}

\usepackage[scale=1., varg]{newpx}
\usepackage{unimath}
\usepackage{preamble}
\usepackage{sidecap}
\usepackage{xspace}
\usepackage{microtype}
\usepackage{natbib} \bibpunct[, ]{(}{)}{,}{a}{}{,}
\graphicspath{{./}{./figsuper}}
\jname{rsta} \Journal{Phil. Trans. R. Soc. A}
\hypersetup{pdfencoding=auto,unicode, psdextra, hidelinks}
\newcommand{\MG}{Matsuno--Gill\xspace}

\begin{document}

%\pagenumbering{arabic} \setcounter{page}{1}

%%%% Article title to be placed here
\title{Jets and Superrotation in Deep and Shallow Planetary Atmospheres}

\author{%%%% Author details
Geoffrey K. Vallis$^{1}$, Loren I. Matilsky$^{2}$ and Quentin Nicolas$^{3}$}

%%%%%%%%% Insert author address here
\address{$^{1}$University of Exeter\\
$^{2}$University of California, Santa Cruz\\
$^{3}$ETH, Zurich}

%%%% Subject entries to be placed here %%%%
\subject{Geophysical Fluid Dynamics}

%%%% Keyword entries to be placed here %%%%
\keywords{superrotation, jets, antidiffusion, planetary atmospheres. geophysical fluid dynamics}

%%%% Insert corresponding author and its email address}
\corres{Geoffrey K. Vallis\\
\email{gkvallis@gmail.com}}

%%%% Abstract text to be placed here %%%%%%%%%%%%
\begin{abstract}
\begin{minipage}{0.55\textwidth}
\small
Zonal flows in planetary atmospheres are ubiquitous, and nearly all the planets in the Solar System have flows that are zonally rather than meridionally aligned.  \textit{Jets,} which are essentially  concentrated streams of fluid that are distinct from a more quiescent background, are less common but can also be found in both deep and shallow atmospheres.  Superrotation,  which in most circumstances  simply means prograde motion (relative to the planetary rotation)  at the equator, is  less common but can  be found in both deep and shallow planetary atmospheres, and in both quickly and slowly-rotating atmospheres: Jupiter, Saturn,  Venus and Titan all have superrotating atmospheres.   Jets, especially superrotating jets, imply some form of  `antidiffusion' of momentum,  meaning that momentum (or angular momentum) must be transported upgradient.   This article discusses some of the mechanisms that give rise to jets and superrotation in both deep and shallow planetary atmospheres, on both slow and fast  rotators, contrasting and comparing the processes involved.  Topics discussed include the roles of convection in deep atmospheres,   geometric and topographic  $\beta$-effects, potential-vorticity homogenization,  wave--mean-flow interaction and tidal locking in exoplanets.
\end{minipage}
\end{abstract}

% Insert the texts which can accomdate on firstpage in the tag "fmtext" %%%%%
%\begin{fmtext}
%Hello
%\end{fmtext}

\maketitle

\newpage

\section{Introduction}
\label{sec:introduction}

Zonal flows are a common feature in planetary atmospheres, at least in our Solar System and most likely  also in exoplanets.  The underlying reason for this is simple:  rotation inhibits strong meridional flows on the surface of a spherical body because if a parcel of fluid were to move any distance meridionally the conservation of angular momentum would lead to strong zonal flows.  Those zonal flows might be unstable, or be associated with strong meridional temperature gradients,  but in any case the zonal flows would soon be stronger and more persistent than the meridional flows.  A similar argument using the vorticity equation on the $\beta$-plane implies that the meridional velocity, $v$, must be small if $\beta$, the latitudinal gradient of the Coriolis parameter $f$,  is large.   Put simply, the barotropic vorticity equation has the form $\beta v = $`other terms',  and if $\beta$ increases while the other terms stay approximately the same, then $v$ must diminish but $u$, the zonal velocity, need not.
 (On the sphere, $f = 2 \Omega \sin \theta$ where $\Omega$ is the planetary rotation rate and $\theta$ is latitude, and $\beta = a^{-1} \partial f/\partial \theta = 2 a^{-1}\Omega \cos \theta$.  On the commonly used $\beta$-plane $f = \f_0 + \beta y$ where $\f_0$ and $\beta$ are constants and $y$ is the distance in the meridional direction.)

Such simple arguments cannot explain how \textit{jets} form, or in particular how superrotation forms.   Jets may be regarded as concentrated streams that are distinct from a more quiescent background. Superrotation is prograde flow (relative to the planetary rotation) for which a ring of fluid  going round the planet has an average angular momentum exceeding that of the planet itself at the equator at some reference level, commonly taken to be at the solid surface for terrestrial planets and at some reference pressure in the case of gas or ice giants. (This can lead to some arbitrariness, especially if one seeks to extend the definition to stars.)  In nearly all cases superrotation is associated with prograde motion at the equator; if the superrotation were only at higher latitudes the flow would be inertially unstable.

Jets and superrotation both require some form of upgradient transfer of momentum:  diffusion will seek to dissipate any extremum of velocity and thus there must be some other process to counter that effect.   Superrotation has been observed on both deep and shallow atmospheres. The former are often associated with giant planets, and the atmospheres of both Jupiter and Saturn superrotate whereas those of Uranus and Neptune do not.  (Henceforth we may say `the planet superrotates', and it should be understood that it is the atmosphere that is superrotating.)  Shallow atmospheres are typical of terrestrial atmospheres and in the Solar System  both Venus and Titan superrotate, whereas on average Earth does not,  although it does from time-to-time, sometimes associated with the quasi-biennial oscillation.   Outside of the Solar System there is compelling evidence, albeit somewhat indirect, that some tidally-locked planets --- both hot Jupiters and super-Earths --- superrotate.

It is the purpose of this article to describe and discuss the various mechanisms that might give rise to jets and superrotation in various forms of planetary atmospheres.   It is not a  literature review;  rather it is a synthesis and focused discussion of the various mechanisms involved in deep and shallow atmospheres on both slowly- and quickly-rotating planets. We seek to draw attention to commonalities and differences,  indicating how robust or delicate various mechanism may be, and what the remaining problems and uncertainties are.  We give  few observational examples: a description of the observations is not our goal and would require an article unto itself. A number of review articles on related topics have previously appeared that cover observations, models or particular cases in more detail, for example \citet{Imamura_etal20}, \citet{Read_Lebonnois18},   \citet{Pierrehumbert_Hammond19}, \citet{Read24},  and some of the articles in the collection by  \citet{Galperin_Read17}.   References to the primary literature are given in main body of the paper.    The first couple of sections below reprise material that can be found in standard texts but which is included because it is foundational and keeps the discussion self-contained. 

\section{Jets and Rossby Waves}
\label{sec:JetsRossby}

The production of jets by the effects of Rossby waves on the mean flow is a well-known mechanism that is responsible for the production of the `eddy-driven' zonal jet in Earth's midlatitudes and that likely occurs on other shallow atmospheres, and perhaps in a different form on deep atmospheres.  

\subsection{An eddy-driven jet}
\label{sec:eddyjet}
A familiar, and probably the best understood, mechanism involving Rossby waves involves the localized generation and meridional propagation of such waves, leading to an eastward jet in the region of the wave generation (\citealp{Thompson71, Thompson80},  \citealt{Dickinson69}). Suppose that Rossby waves are excited within some compact range of latitudes and that the dissipation mechanisms  or refraction are insufficient to prevent the meridional propagation of the waves away from the source region.  In incompressible two-dimensional flow on the $\beta$-plane the familiar dispersion relation for Rossby waves is
\begin{equation}
	\label{jr:1} 
   	\omega = U k -  {\beta k \over {k^2 + l^2}  }, 
\end{equation}
where $\omega$ is the frequency,  $k$ and $l$ are zonal and meridional wavenumbers respectively, $U$ is the (here assumed unsheared) constant background zonal flow, and $\beta$ is a constant, as introduced earlier. 

\begin{figure}
%\hspace{-0.5cm}
\centering
 \includegraphics[ height=0.35\textwidth]{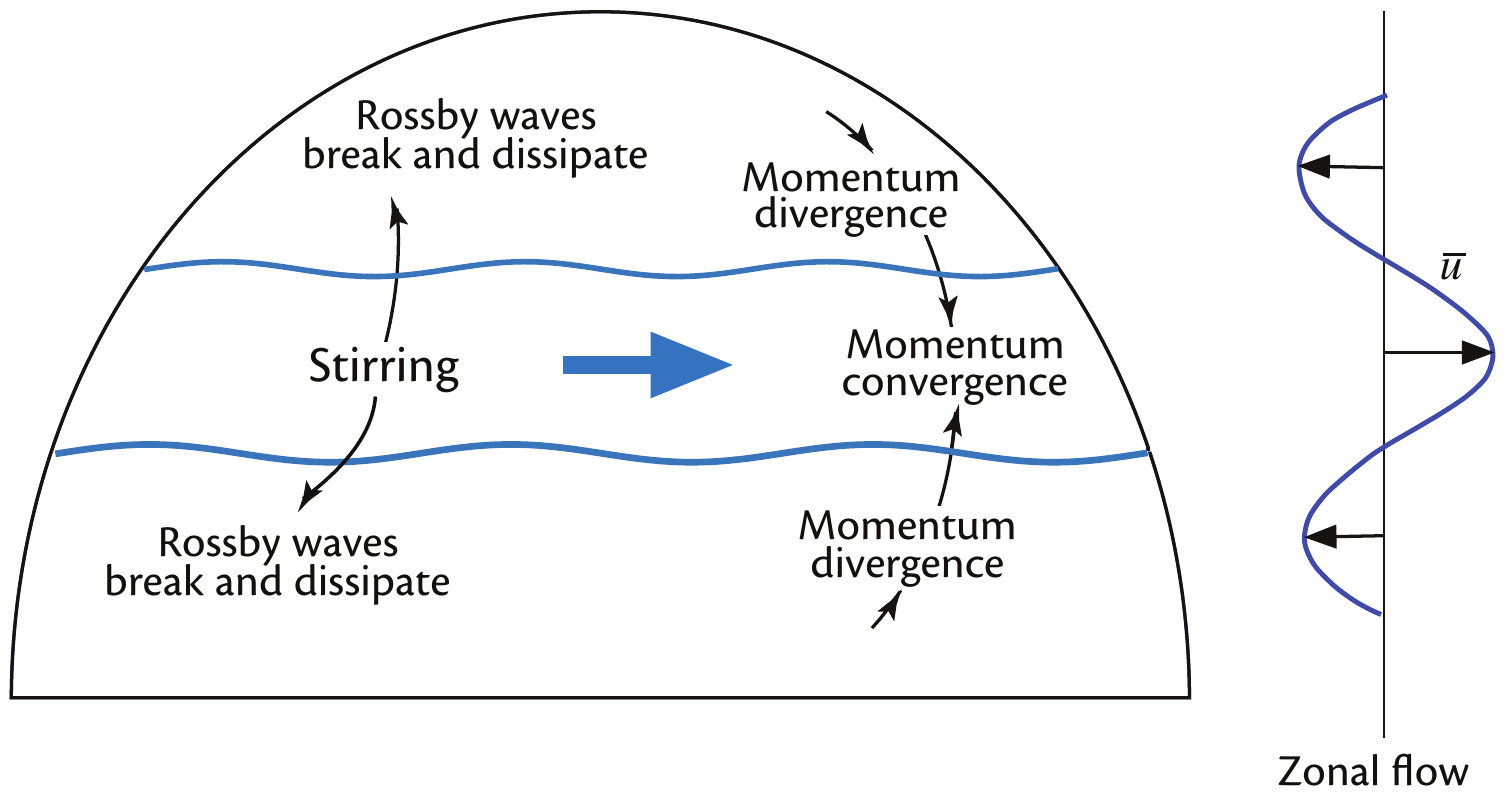}%
\includegraphics[height=0.34\textwidth]{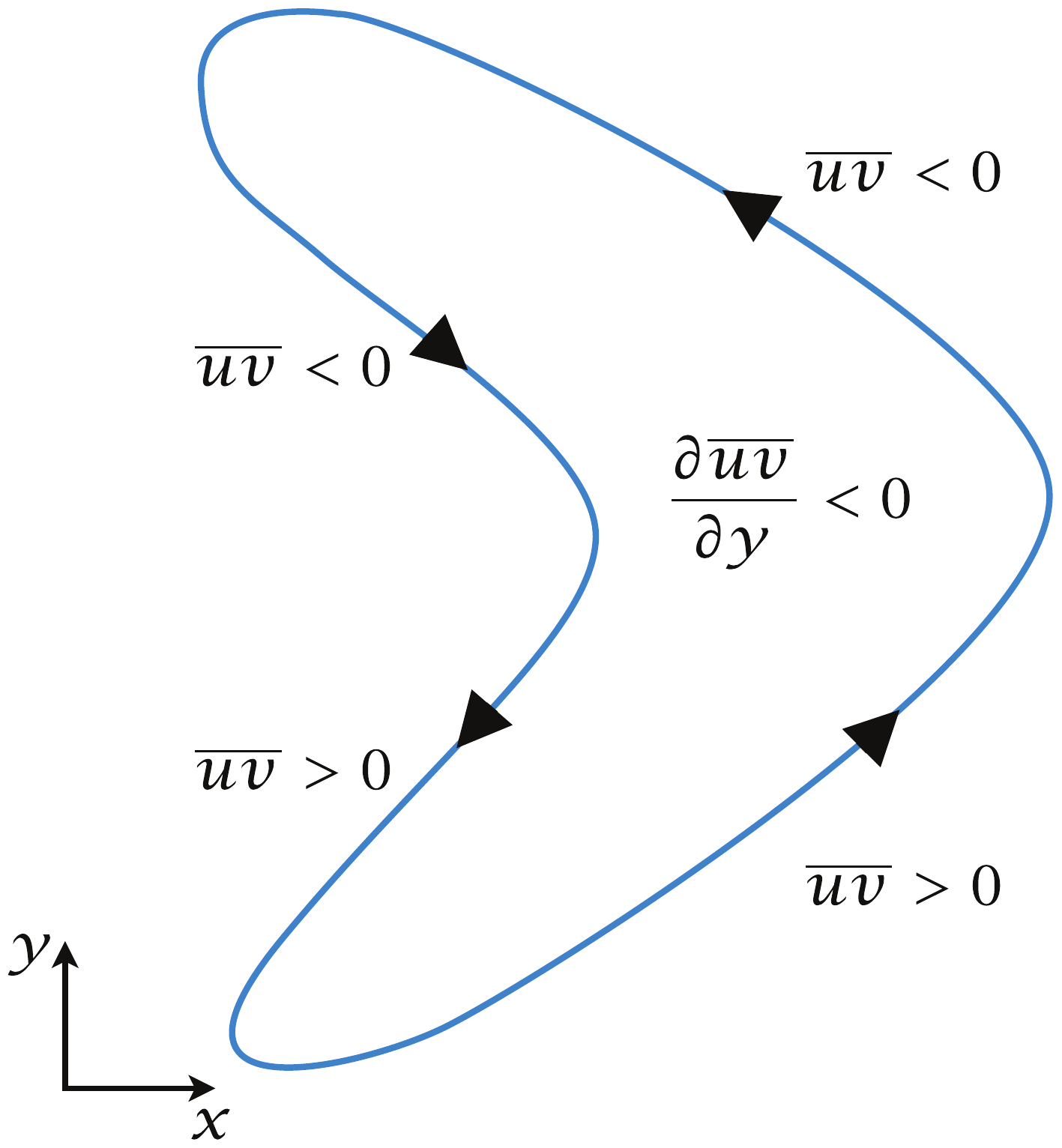}
\caption {Generation of zonal flow on a rotating sphere or $\beta$-plane.
Stirring in mid-latitudes (for example by baroclinic eddies) generates Rossby waves
that propagate away. Momentum converges in the region of stirring, producing eastward flow there and weaker westward flow on its flanks. The right-hand plot shows the momentum transport in physical space produced by the boomerang-shaped eddies produced by the waves. Adapted from \citet{Vallis17}.}
\label{fig:eddyforc}
\end{figure}

The meridional component of group velocity is 
\begin{equation} \label{jr:2}
c_{g}^y = \pp \omega {l} = {2 \beta k l \over (k^2 + l^2)^2 } ~ .
\end{equation}
and this must be directed away from the source region, since energy is carried with the group velocity.  For incompressible flow, the velocity components in the zonal ($x$)  and meridional ($y$)  directions  associated with the Rossby waves are
\begin{equation} \label{jr:3} \eqnab
u'  = - \Re C \, \ir l  \e^{\ir (k x + l y - \omega t)}, \qquad
v'  = \Re C \, \ir k  \e^{\ir (k x + l y - \omega t)} ,
\end{equation}
where $C$ is the complex amplitude of the streamfunction of  the waves. 
The momentum flux  (taking density equal to one) carried by the waves is therefore
\begin{equation}  \label{jr:3b}
\obar{u' v'} = -\half |C|^2 k l ,
\end{equation}
with the overbar denoting a zonal average and primes a deviation from the average. 
The momentum flux associated with the Rossby waves thus has the \textit{opposite} sign to that of the group velocity;  thus, if Rossby waves travel away from the source region, momentum fluxes toward the source region, converging in that region and generating a prograde flow, as in \figref{fig:eddyforc}.  The mechanism relies in a fundamental way on the presence of a potential-vorticity gradient that allows Rossby waves to propagate.

 The mechanism may also be succinctly described by the Transformed Eulerian Mean (\TEM) formalism;  this can lead to some advantages in more complicated situations, for example stratified flow involving buoyancy fluxes in which use of the potential vorticity and/or Eliassen--Palm fluxes can be helpful.  In any case, it is the presence of a potential vorticity gradient allowing for the propagation of Rossby-like waves that is essential, and this can be produced  by differential rotation, the effects of stratification, or topographic effects; in all cases the essential mechanism is that above, or variations on that effect \citep{Vallis17}. 
 
  It is this mechanism that produces an eastward, largely barotropic flow, and an associated eastward surface flow, in Earth's midlatitudes, sometimes called the eddy-driven jet. This jet should be mechanistically distinguished from the general eastward thermal wind that is associated with the pole–equator temperature gradient, although this too produces a prograde flow when high latitudes are cooler than low ones \citep{Marshall_Plumb08, Vallis19}.  On Earth there is subtropical jet at the edge of the Hadley Cell that has very little eddy forcing,  being mainly comprised of a vertically sheared zonal flow in thermal wind balance.  In some contrast, the midlatitude jet has an eddy-driven component with a strong barotropic component.  This jet is superimposed on the thermal wind, and the subtropical and eddy-driven jets are often merged and observationally barely distinguishable from each other even if the mechanisms producing them are distinct .   In the upper troposphere on Earth the broad eastward flow that arises from the thermal wind balance is at least as large as the eddy-driven component,  but this is not obviously the case on the giant planets. 

\subsection{Multiple jets on the $\beta$-plane and sphere} \label{sec:mutiple}

Multiple, alternating east-west,  jets can also form on the \bta-plane or on the surface of a rotating sphere. However, the description of  their formation commonly differs from that above, being couched in the language of two-dimensional (or geostrophic) turbulence in which the governing equation is the barotropic vorticity equation,
\begin{equation}
	\label{baro:1} 
   	\pp \zeta t + J (\psi, \zeta) + \beta \pp \psi x = - r \zeta + \nu \nabla^2 \zeta + F. 
\end{equation} 
Here, $\psi$ is the streamfunction of the flow (such that $u=-\partial_y\psi v = \partial_x\psi$), $\zeta = \nabla^2 \psi$ is the relative vorticity, $J$ is the Jacobian ($J(\psi, \zeta) = \pppp \psi x \pppp \zeta y - \pppp \psi y \pppp \zeta x$), $r$ and $\nu$ are linear drag and viscous coefficients (respectively) and $F$ is the imposed forcing, often supposed to be at some small scale.  Energy is then generated at small scales (where the nonlinear terms dominate) and is transferred by some form of inverse cascade to  large scales where flow becomes more linear and Rossby waves dominate, leading to jet formation.  A scaling for the crossover scale, simply based on dimensional arguments, is 
\begin{equation}
	\label{baro:2} 
   	L_R  = \bfracsup{U}{\beta}{1/2},
\end{equation}
where $U$ is some representative  fluid speed of the eddying flow, and this scale is commonly called, and taken to be the definition of,  the Rhines scale \citep{Rhines75}.   Because the  estimate is obtained by comparing a turbulence speed to a Rossby wave speed, $U$ is most naturally thought of as a root-mean-square (rms) velocity  --- ideally at the crossover scale itself --- and  not the velocity of the jet.  In some circumstances the mean zonal velocity itself might be used as a rough measure of $U$ but this is not generally appropriate. 

Equation \eqref{baro:2} is not a prediction for the crossover scale, since $U$ is not known a priori nor is it always easily measured.  Instead, one may take the energy injection rate, $\epsilon$ (which is equal to the cascade rate), to be known for this is an external parameter.  In numerical simulations one can control the value of $\eps$, and in planetary atmospheres one may be able to determine it if the internal heat flux is known and/or from the radiation budget at the top of the atmosphere.    Given $\eps$,  a standard turbulence-based scaling connecting the velocity $U$ with scale $L$ is
$\eps \sim (U^3  / L)$.  Using this in \eqref{baro:2}  gives the $\beta$-scale \citep{Vallis_Maltrud93},
\begin{equation}
	\label{baro:2b} 
   	L_\beta = \bfrac{\eps}{\beta^3}^{1/5} ,
\end{equation}
This scale is not  the scale at which the inverse cascade ceases; rather, it is the scale at which the flow becomes more zonally oriented through the action of Rossby waves that allow the energy to proceed to large scales only if it becomes zonally elongated, and so become jet-like \citep{Vallis_Maltrud93}.  (Wave-wave interactions among Rossby waves themselves also have a tendency to produce zonal flows \citep{Newell69, Rhines75}.) The inverse cascade finally halts due to the effect of linear drag, and if the rms velocity  at the halting scale is given by  $U \sim (\eps/r)^{1/2}$ then, using this estimate in \eqref{baro:2}, we obtain the scale
\begin{equation}
	\label{baro:2c} 
   	L_{\beta r} = \bfrac{\eps}{r \beta^2}^{1/4} .
\end{equation}
This scale  is not the only one that can be constructed form $\beta, \eps$ and $r$  but it is  physically motivated, and for small friction $L_\beta < L_{\beta r}$.   The length scales given by \eqref{baro:2b} and \eqref{baro:2c} may be put in the form of $(U/\beta)^{1/2}$, with particular choices of $U$, and in that sense are Rhines scales, but for clarity are often referred to as the $\beta$-scale and $\beta$-$r$-scale respectively.   The scale around $L_{\beta r}$, if larger than $L_\beta$, does seem to be the meridional scale at which zonal jets form in numerical simulations \citep{Galperin_etal04}, and so is a measure of their width.   The region between $L_\beta$ and $L_{\beta r}$ (when the latter is larger) is known as the zonostrophic regime and scaling arguments suggest that the kinetic energy spectrum  in that region is anisotropic and should follow $E(k) = C_z \beta^2 k^{-5}$ for small zonal wavenumber, where $k$ is the total wavenumber and  $C_z$ is some constant \citep{Sukoriansky_etal02}.    

The case when $r$ tends to zero is an interesting one, whether or not $\beta$ is non-zero.  If the Laplacian viscosity is also small then, in the barotropic (two-dimensional) case there is no mechanism to halt the inverse transfer of energy. The scale of motion, be it zonal flows or vortices, will continue to grow, limited only by the size of the domain,  with the kinetic energy continuing to pile up at that scale \citep{Kraichnan67}.  A Laplacian viscosity cannot prevent this  unless it is large enough to operate across all scales,  in which case the flow is no longer turbulent.   In a baroclinic fluid (and so in planetary atmospheres) there is another escape route for the kinetic energy, namely conversion to baroclinic energy at large scales;  the energy there can be thermally damped, and this may allow the fluid to equilibrate even with zero drag \citep{Chai_etal16}.  This route may be important in giant planets although it has yet to be fully explored. 

The mechanism of multiple jet formation should not be regarded as wholly different from that described in \secref{sec:eddyjet}, in that  each prograde jet is still proximately produced by eddy momentum flux convergence from Rossby wave propagation.  But in the homogeneous, multi-jet case the position of each jet is more arbitrary and not necessarily determined by the position of some localized external forcing.

\subsection{Potential vorticity homogenization and staircases}
\begin{SCfigure}
\hspace{-0.6cm}
 \includegraphics[ width=0.6\textwidth]{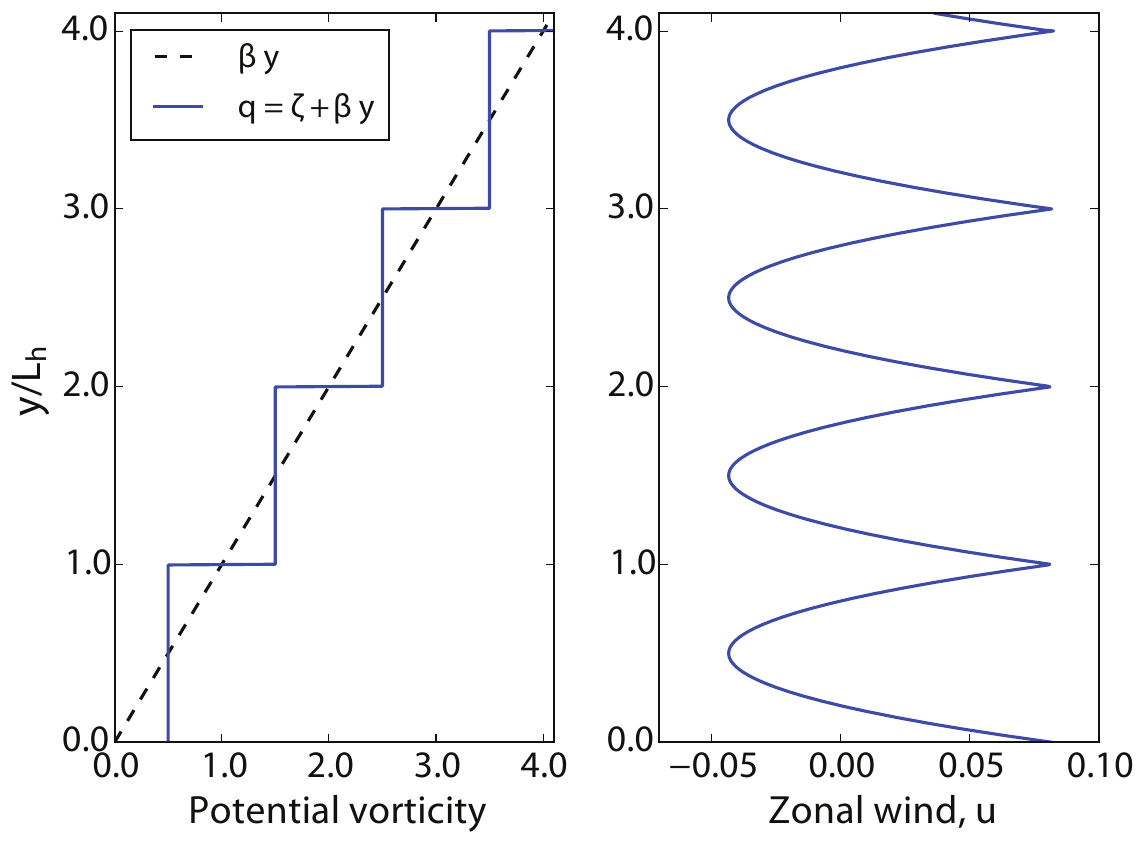}%
   \caption{An idealized potential vorticity staircase. The left panel shows homogeneous regions of potential vorticity separated by jumps, with the homogenization occurring over a scale $(U/\beta)^{1/2}$. The right-hand panel shows the corresponding zonal flow. Adapted from \citet{Vallis19}.}
\label{fig:stair}
\end{SCfigure}

We may write \eqref{baro:1} in the equivalent form
\begin{equation}
	\label{baro:3} 
   	\DD q  = F - D. 
\end{equation} 
where $q = \zeta + \beta y$ and $F$ and $D$ represent forcing and dissipation; evidently, the  above equation has the same form as that governing a passive scalar.  If the flow is turbulent but the forcing and dissipation are weak there will be a tendency for the potential  vorticity to homogenize \citep{Rhines_Young82b},  with gradients of potential vorticity being expelled to a boundary. (This result is a variation on the well-known Prandtl--Batchelor effect, that vorticity in a region surrounded by a closed streamline must be constant.) However,  there may be insufficient energy for the entire domain to become homogenized, in which case regions may become homogeneous with rapid gradients between them and in a zonally-re-entrant domain  a potential vorticity staircase may form (\figref{fig:stair}) with most of the kinetic energy in the zonal flow.   \citet{Marcus93} and  \citet{Marcus_Shetty11} proposed the relevance of this mechanism to jet formation in planetary atmospheres and others have explored the mechanism more generally \citep[e.g.,][]{Dritschel_McIntyre08, Scott_Dritschel12}.   A closely related mechanism producing staircase-like structures of density in stratified flow was previously identified in an oceanographic context by \citet{Phillips72} and \citet{Posmentier77}, and the analogy to potential vorticity mixing and inverse cascade was further cemented by \citet{Cabanes_etal20} using the sorting algorithm of \citet{Thorpe05}.   Earlier observations of layering were actually made by  \citet{Tait_Howe68} and  thermohaline staircases in stratified fluids  are now fairly commonly observed.

\subsection{Deep and shallow atmospheres} \label{sec:deepshallow}
The mechanisms of jet formation described above are two-dimensional; that is to say, they can be described by equations in which the fields vary only in latitude and longitude.   In a shallow atmosphere in which the atmospheric depth is much less than the planetary radius, which is typical of the terrestrial planets in the Solar System, quasi-two-dimensionality can be regarded as a reasonable approximation:  indeed the quasi-geostrophic equations (which do allow for vertical structure) exhibit similar behavior to the purely two-dimensional equations, for they have a conserved potential vorticity that is more-or-less a Laplacian of the stream function, an inverse cascade of energy that can be diverted by a $\beta$-effect,  and a forward cascade of enstrophy \citep{Charney71, Salmon80}.  However, the gas giants do not have shallow atmospheres -- the planets are themselves fluids, albeit not always gaseous and with densities that increase enormously with depth. That said, there are two effects that may make the atmospheres behave in a quasi-two dimensional fashion, as follows.

\begin{figure}
    \centering
    \includegraphics[width=0.75\textwidth]{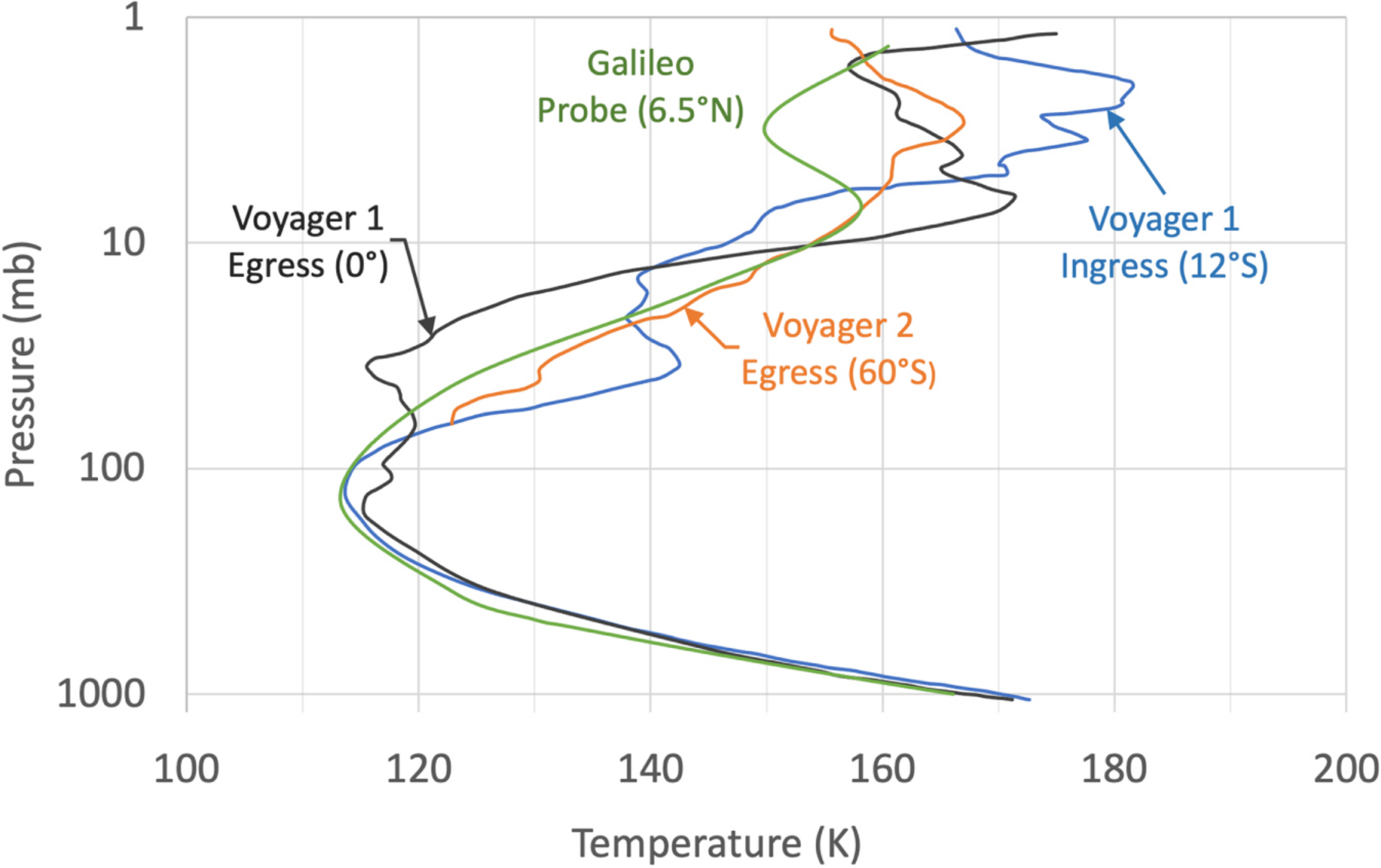}
       \includegraphics[width=0.75\textwidth]{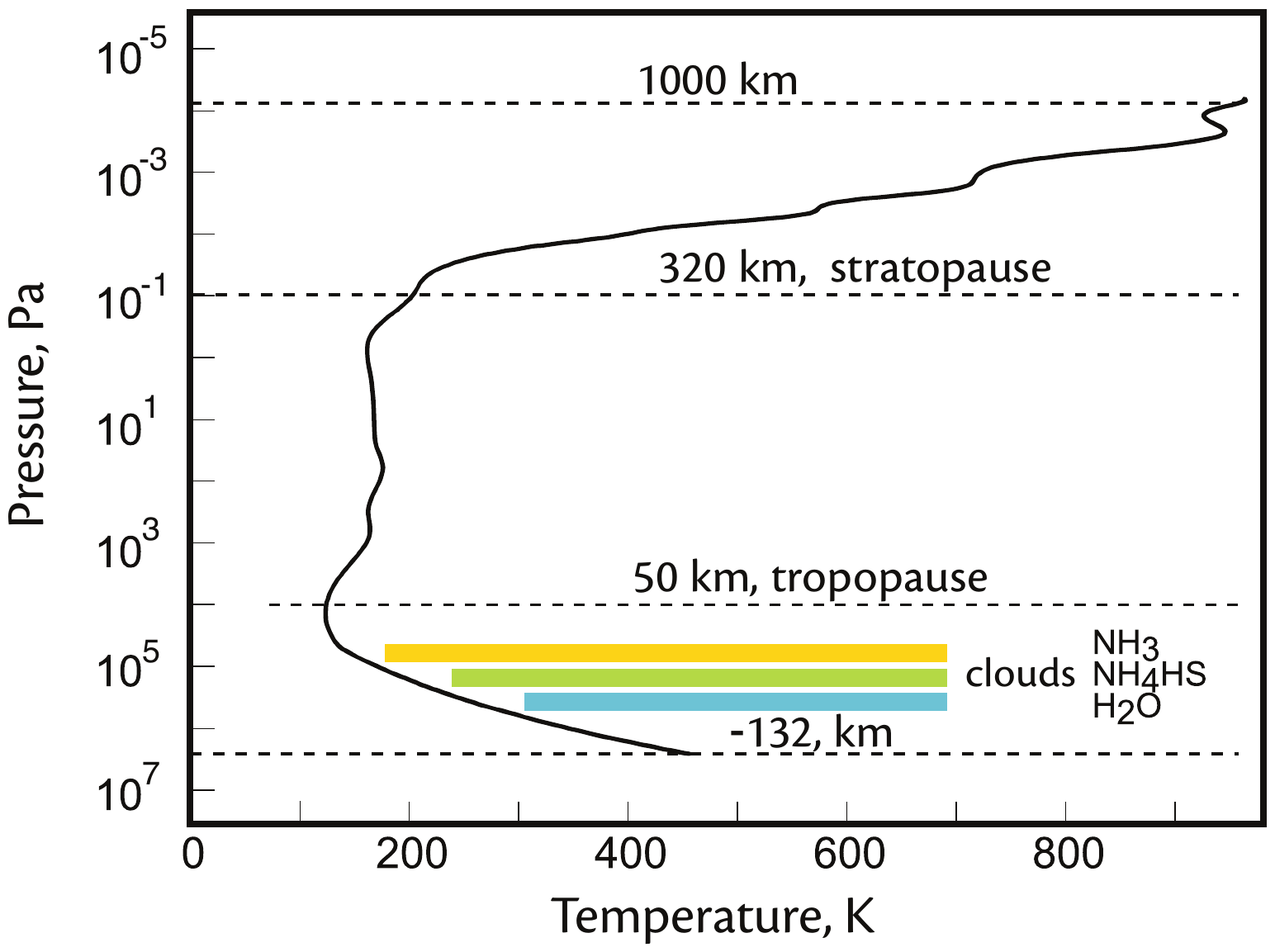}
    \caption{Temperature in Jupiter’s upper atmosphere.   Panel (a), from \citet{Gupta_etal22}, uses data from Voyager radio measurements.  Panel (b) is adapted from Ruslik0 at en.wikipedia using data from the Galileo probe, with heights measured from the $10^5$ Pa (1000 mb)  level. }
    \label{fig:Jupiterstrat}
\end{figure} 

The first is simply that the upper atmospheres of the gas giants (the `weather layers') are, at least to some degree, stably stratified  (\figref{fig:Jupiterstrat}). Although the most direct evidence for this comes from a single location using the Galileo probe, temperature profiles obtained from Voyager occultation measurements (\citealt{Lindal_etal81}, \citealt{Gupta_etal22}) give a qualitatively consistent view.  Stratification may arise through radiative effects (with radiation becoming more efficient than convection in transporting energy upward in the weather layers)  and/or through moist convection;  which of the two is the dominant mechanism (if there is one) in the Solar System gas giants is not known.  Stratification inhibits vertical motion and leads to a more two-dimensional flow,  and if the Rossby number is low the combined effects lead to the quasi-geostrophic equations. 

The second reason is that,  even if the flow is deep, the presence of rapid rotation 
organizes the flow into columnar structures which are largely invariant along the axis of rotation---a manifestation of the Taylor--Proudman effect.   This effect has manifestations even when the flow is convecting, enabling the flow to be described by a set of quasi-geostrophic-like equations \citep{Busse70, Busse76, Julien_etal06} that have similar properties to those of the more familiar quasi-geostrophic equations for shallow, stratified flow. 

\begin{figure}
    \centering
    \includegraphics[height=0.45\textwidth]{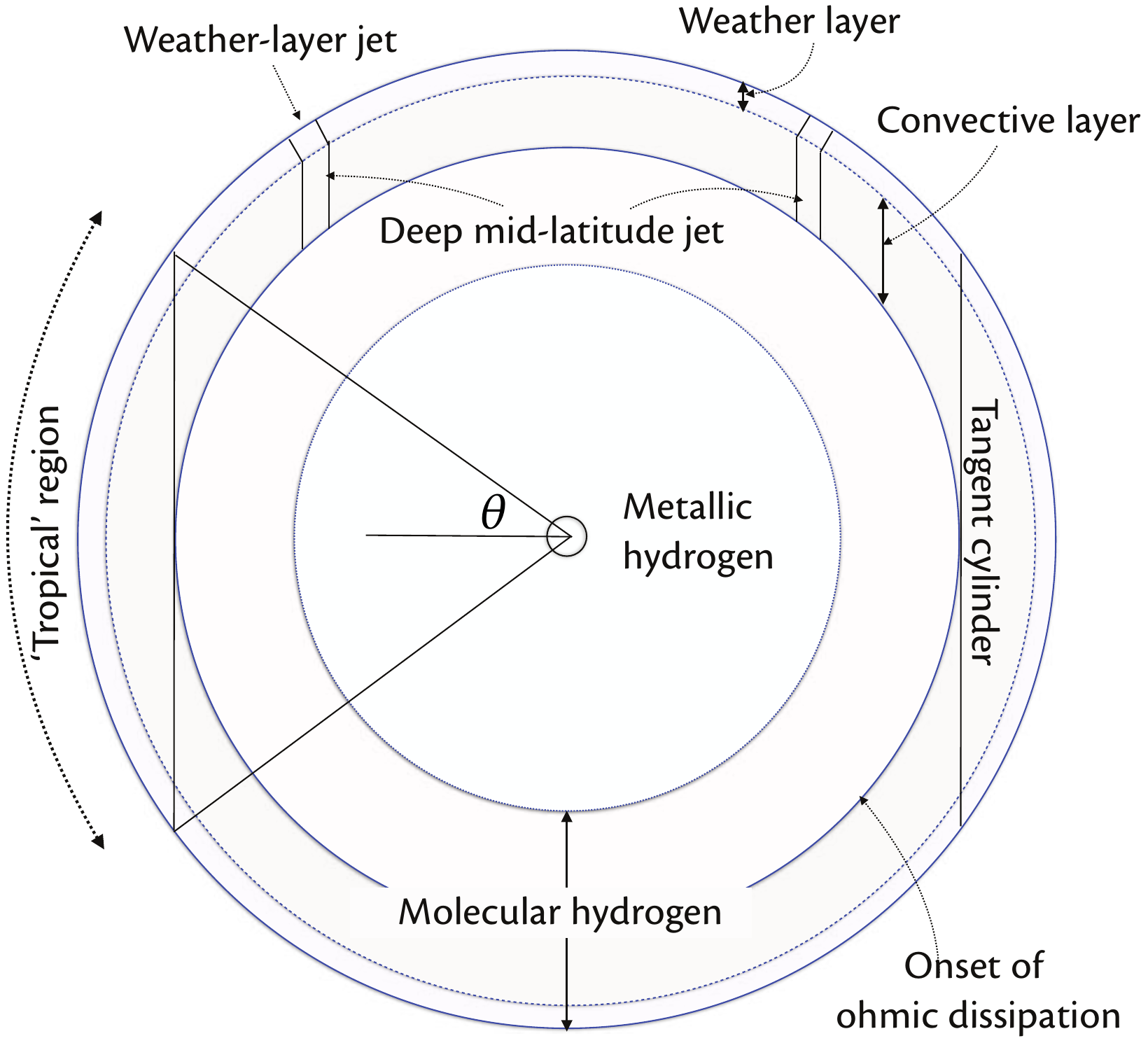} \hspace{0.02\textwidth}
    \includegraphics[height=0.45\textwidth]{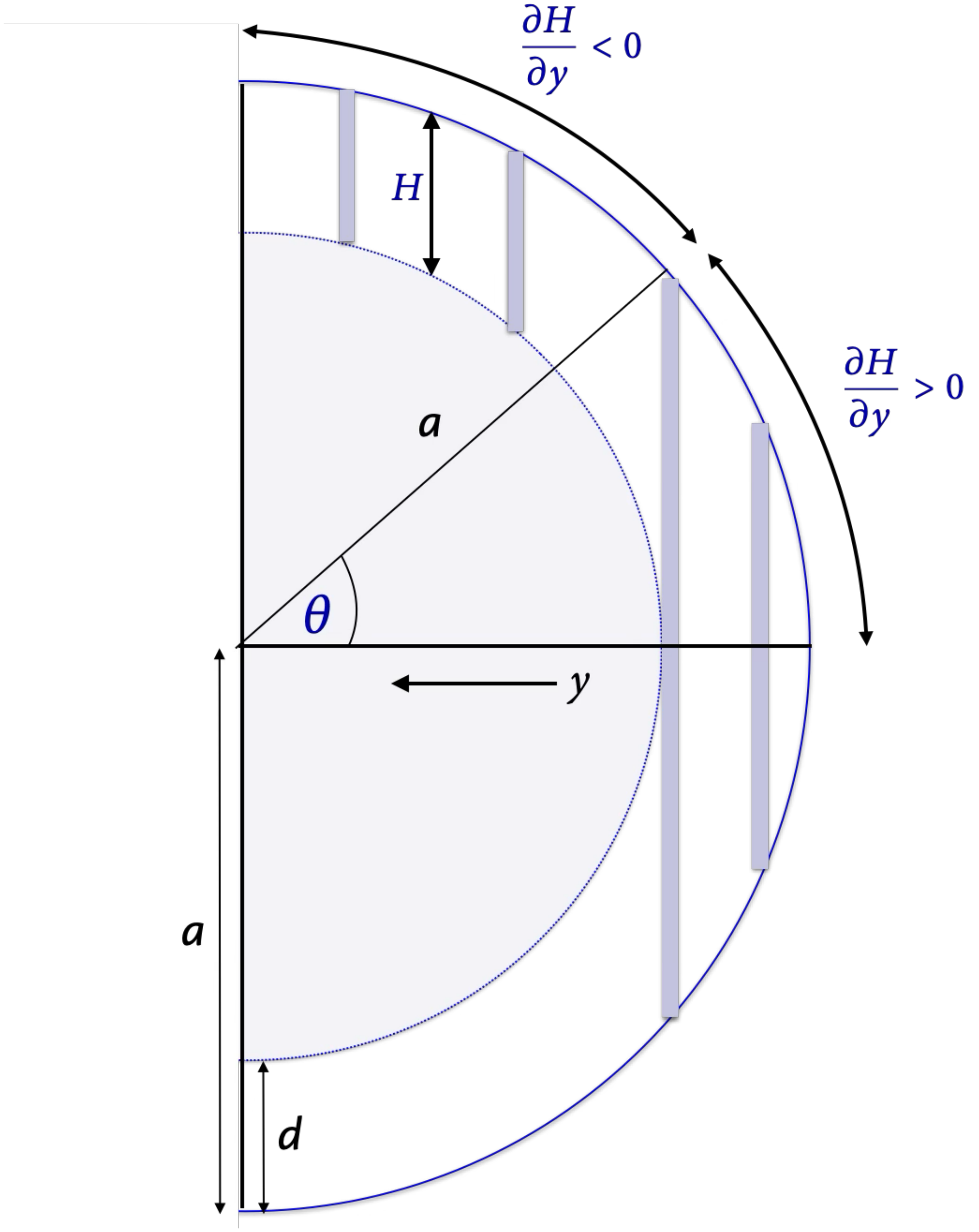} 
    \caption{idealized structure and geometry of a giant planet such as Jupiter or Saturn.  The right-hand panel shows how the thickness of the convecting shell changes with distance from the rotation axis, and so with latitude.  Adapted from \citet{Vallis19}.}
    \label{fig:geometry}
\end{figure}

The above two effects have led to competing views of the origin of the jets observed on the giant planets in the Solar System.  One mechanism emphasizes  the weather layer, in which geostrophic turbulence on the $\beta$-plane can lead to jet formation through the mechanisms described in Sections 2\ref{sec:eddyjet} and 2\ref{sec:mutiple}.  A mechanism for superrotation is less obvious in this context, although if Rossby waves can be generated at the equator and propagate poleward in each hemisphere then equatorial superrotation can arise (\citealp{Schneider_Liu09}, \citealt{Lian_Showman10}).  In contrast, Busse suggested that  the jets on Jupiter are a manifestation of deep convection with the role of any weather layer being secondary. The convective cells become aligned with the axis of rotation and a topographic $\beta$-effect arises because the depth of the shell in the direction of the axis of rotation varies with cylindrical distance from the axis of rotation, and so with latitude (\figref{fig:geometry}). The flow can then take on a quasi-geostrophic form and zonal jets can form.  In the equatorial region the variation of shell depth is the opposite to that in mid-latitudes (\figref{fig:geometry}) so that an effective beta,  now due  to  variations in column depth, is \textit{negative}. This gives a natural mechanism for the formation of superrotation, as we discuss in Section 3\ref{sec:deep}. 

Gravimetric observations from the Juno mission \citep{Kaspi_etal18} indicate that the jets on Jupiter \textit{do} extend into the atmosphere, probably to a depth of order 3000 km and likely with some variation in latitude, with those on Saturn extending still deeper \citep{Galanti_etal19}. The Jovian jets also appear to be aligned with the axis of rotation, at least below the weather layer.   Although a few thousand kilometers  is not deep compared to the radius of these planets it is deep enough for a full treatment of the Coriolis effect to be needed. The low density of fluid in the weather layer, and the relative shallowness of that layer (about 100 km), mean that the gravitational influence of the layer is too small for gravimetric observations to discern whether the flow is aligned along the axis of rotation or along radial lines  in that region.

\section{Mechanisms of Superrotation}
\label{sec:mechanisms}

We now turn our attention to the mechanisms producing superrotation, and in particular prograde motion at the equator.

\subsection{Constraints on Superrotation and Hide's Result} \label{sec:Hide}

As noted, superrotation may be defined as the presence of a zonal flow whose average angular momentum around the planetary axis of rotation exceeds that of the  planet itself at some reference level (e.g., the planet surface) at the equator.  For simplicity, consider a shallow atmosphere (meaning its depth is much smaller than the planetary radius, $a$) on a terrestrial planet. If the planet rotates at a rate $\Omega$ and the zonally-averaged zonal velocity is $\ubar$, then the average angular momentum (per unit mass) of a ring of fluid at a given $\vartheta$ is given by 
\begin{equation}
    \label{H:1}
    m = (\Omega a \cos\vrt    + \ubar) a \cos \vrt. 
\end{equation}
Let us suppose that $\ubar = 0$ at the equator (and thus $m(\vrt=0) = \Omega a^2$). A ring of air at the equator that moves poleward will, in the absence of viscosity, conserve its angular momentum so that at a latitude $\vrt$ its zonal wind is given by
\begin{equation}
	\label{H:2} 
   	\ubar(\vrt) = \Omega a {\sin^2\vrt \over \cos \vrt} .
\end{equation}
If its velocity exceeds this, the ring will be superrotating.  However, this is not easily achieved for the simple reason that $m$ \textit{is} conserved.  That is, it obeys an equation of the form
\begin{equation}
	\label{H:3} 
   	\DD m = D + E
\end{equation}
where $D$ represents dissipation and $E$ represents eddy effects. If $E=0$ and $D$ is due to Laplacian viscosity and if the vertical scales are much smaller than the horizontal scales, then the viscous term is  $\nu \ppp[2] m z$. In this case dissipation can only dissipate extrema of angular momentum and no superrotation can form, and this is essentially Hide's result \citep{Hide69}.   (If viscosity is large enough to be dominant the flow will tend toward three-dimensional solid-body rotation, but this is very unlikely if the dissipation is due to molecular effects.)   However, superrotation can potentially form if the  eddy terms, $E$,  in \eqref{H:3} are non-zero, and \citet{Gierasch75} suggested that vorticity mixing could give rise to superrotation. Various other forms of eddy viscosity and drag have also been proposed \citep[e.g.,][]{Sukoriansky_etal99}.   In the next sections  we focus on the mechanisms of how eddies and/or wave--mean-flow interaction gives rise to superrotation.

\subsection{Superrotation in Terrestrial Atmospheres}
\label{sec:terrestrial}

Two rather different kinds of terrestrial planetary bodies are known to superrotate: slowly-rotating  planetary bodies (such as Venus and Titan) and some tidally-locked planets.  A robust mechanism producing superrotation on both classes of planets involves the interaction of Rossby and Kelvin waves, although the forcings of these waves differs considerably  in the two cases.  (Venus  also has very long solar days and tidal forcing is likely to be important, and many tidally-locked planets are gas giants, effects we do not consider here.)

\subsubsection{Slow rotators}

\begin{figure}
    \centering
    \includegraphics[width=\textwidth]{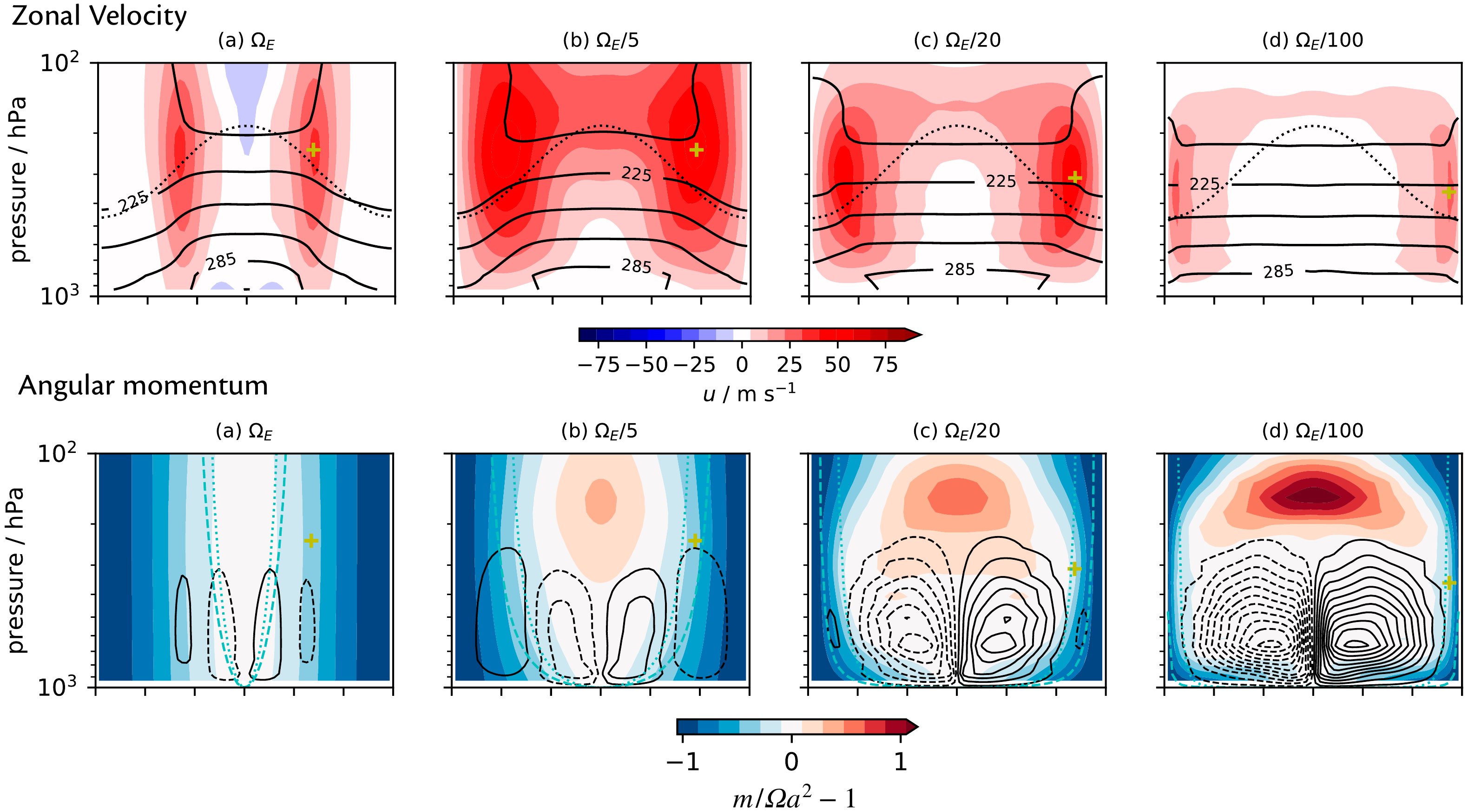}
    \caption{A series of numerical integrations using an idealized GCM \citep[Isca, ][]{Vallis_etal18} configured by G. Colyer  in an Earth-like setting  but with the rotation rate varying from that for Earth ($\Omega_E$) to $\Omega_E/100$. The upper row shows zonally averaged zonal velocity (colors) and temperature (contours) and the lower panel shows a normalized excess angular momentum (colors), which is a useful measure of superrotation, and meridional overturning streamfunction (contours).  The horizontal axis is latitude, from $-90\textdegree$ to $+90\textdegree$}
    \label{fig:gsuper}
\end{figure}

Numerical simulations suggest that superrotation on slowly-rotating, shallow, planetary atmospheres (henceforth just `slow rotators')  is a robust effect, as illustrated in \figref{fig:gsuper}.  
One candidate mechanism for such superrotation is the production of Rossby waves  in equatorial regions with subsequent meridional propagation, thus converging momentum to the equator in the manner similar to that described in Section 2\ref{sec:eddyjet}. However, on slow rotators a mechanism for the generation of such Rossby waves is hard to envision, and in any case ageostrophic effects and Kelvin waves are likely to be important. Rather, the superrotation mechanism (at least that arising in numerical simulations) is that of an  instability involving the interaction of off-equatorial Rossby waves and equatorial Kelvin waves. This `RK instability'  was  identified by \citet{Iga_Matsuda05}  and further explored  by \citet{Wang_Mitchell14}, and  (now with hindsight) is recognized as the mechanism appearing in \citet{Mitchell_Vallis10} and other simulations since, with \citet{Potter_etal14}  emphasizing the importance of Kelvin waves and  \citet{Zurita-Gotor_Held18} making explicit the connection to RK instability. 

Kelvin waves typically have an eastward phase speed whereas Rossby waves travel westward,  and Kelvin waves decay away from the equator with Rossby waves more dominant at higher latitudes. Given these remarks, two conditions must be satisfied for a significant interaction to occur: (i) The Rossby wave needs to be Doppler shifted eastward by a background flow so its phase speed and frequency match that of the Kelvin wave. (Such an eastward shift can naturally arise if there is an eastward thermal wind in the extratropics. A westward shift of the Kelvin wave is also  possible in some circumstances \citealt{Nicolas_Vallis26}.)  (ii) The Rossby wave and Kelvin wave should have some physical proximity, otherwise the interaction will be weak.  In practice this means the Rossby waves should be generated within a distance of the order of an equatorial deformation radius from the equator. A third, more subtle, requirement arises because Kelvin waves are symmetric across the equator. (Anti-symmetric modes such as Yanai waves do exist but are less easily excited.) If the RK mode is equatorially symmetric (with $v=0$ at the equator at all times) then it cannot converge momentum  to the equator in the absence of vertical momentum transport.  This is because the inviscid zonal momentum equation may be written (using  pressure co-ordinates for convenience) as
\begin{equation}
	\label{sr:0} 
   	\pp ut - (f + \zeta)\, v + \omega \pp u p = - \pp{} x  \Big(  \phi +  {1 \over 2} \ub^2 \Big) .
\end{equation}
where $p$ is the pressure, $\omega$ is the pressure vertical velocity and $\phi$ the geopotential. The zonally-averaged equation is then
\begin{equation}
	\label{sr:1} 
   	\pp \ubar t = \obar {v (f + \zeta)} - \obar {\omega \pppp {u} p } .
\end{equation}
 In a model with no vertical variation the terms involving $\omega$  disappear and the terms involving $\vbar$ alone cannot give rise to superrotation (as discussed in Section 3\ref{sec:Hide}). Omitting them we are left with
 \begin{equation}
 	\label{sr:2} 
    	\pp \ubar t =   \obar {v' \zeta'}  .
 \end{equation}
The term on the right-hand side vanishes if  $v'=0$ at all times.   In a single-layer shallow-water model an essentially similar argument  begins by writing the momentum equation as
\begin{equation}
	\label{sr:2b} 
   	\pp u t  - (f + \zeta)\,v   =  - \pp{}x \Big(\phi + {1\over 2} \ub^2 \Big)
\end{equation}  
After zonally-averaging \eqref{sr:2} is again found.  The point to be made is that $\obar{v' \zeta'}$, rather than $\ppp {\obar {u'v'}} y$ alone, is the relevant eddy momentum flux, and so that if $v'=0$  single-layer models cannot produce superrotation unless vertical momentum transport is parameterized, as inferred  by \citet{Showman_Polvani10}. However,  superrotation in slow rotators is not well reproduced in a  `one-and-a-half' layer model (i.e., a model with a single moving layer overlying a quiescent layer) even when  vertical transport is parameterized \citep{Zurita-Gotor_Held18},  and more vertical structure is needed to properly capture it.

\begin{figure*}[!t]
\begin{center}
    \includegraphics[width=0.9\textwidth]{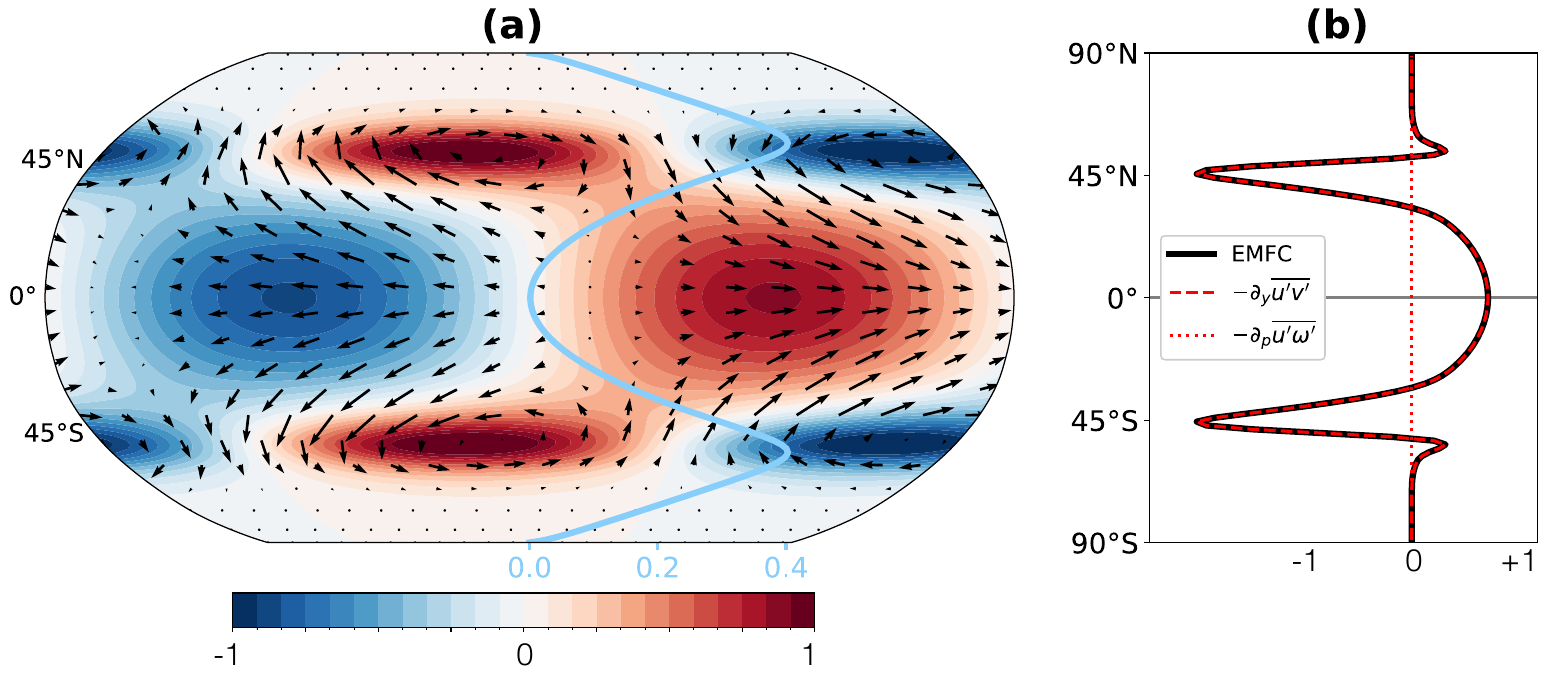}
 \caption{(a) Structure of a Rossby--Kelvin eigenmode in the upper level of a two-level model. Shading shows the geopotential height perturbation and arrows show the horizontal wind perturbation. This is the most unstable mode arising from the linearization of the primitive equations with a barotropic background state whose zonal wind is shown as a blue line.  The bottom light blue scale shows the local Rossby number $U/(2a\Omega)$ of that zonal wind, and the amplitude of the perturbation shown in the color bar is in arbitrary units. The mode propagates eastward at a speed of $0.4 a\Omega$. (b) Eddy momentum flux convergence (black), and its horizontal (dashed red) and vertical (dotted red) components.  Figure adapted from \citet{Nicolas_Vallis26}.}
 \label{fig:RK1}
\end{center}
\end{figure*}

The RK instability produces a characteristic pattern that transports zonal momentum equatorward,  producing a prograde jet straddling the equator.  A sample pattern --- an eigenmode of the instability --- is illustrated in \figref{fig:RK1}, using a two-level primitive equation model, which is perhaps the simplest model that can naturally produce superrotation  by this mechanism.  The basic state is zonally-symmetric, with a pole--equator temperature gradient producing an eastward flow in midlatitudes (the pale blue line in the figure), Doppler shifting the Rossby waves. 
The eddy momentum flux produced by the pattern is shown in the right-hand panel, and is such as to produce a prograde equatorial jet.    Although it appears from the figure that the horizontal momentum flux convergence is solely responsible for the eddy flux convergence, this is slightly misleading.  The eddy flux convergence may also be written as $\obar{v'\zeta'} - \obar{w'\pppp {u' }p}$ and, for this kind of eigenmode $v' = 0$ at the equator.  Thus, at the equator, the eddy flux convergence involves a correlation between the \textit{vertical} velocity and the vertical shear of the zonal flow, emphasizing the importance of the phase difference, or baroclinicity, in the vertical structure of the zonal flows. 

In fully nonlinear integrations, a basic-state zonal wind profile such as that of \figref{fig:RK1} is not externally imposed but naturally appears as a consequence of thermal-wind balance, modified by the action of baroclinic eddies. Slower planetary rotation (or large thermal Rossby number) tends to favor both a wide meridional extent of Kelvin waves and stronger local Rossby number at midlatitudes. Both effects favor RK instability, and indeed superrotation robustly increases as planetary rotation decreases in such simulations (the parameter dependence is explored in \citealt{Nicolas_Vallis26}).   The importance of nonlinear effects was also noted by \citet{Zurita-Gotor_Held18} and, in a  different context, by \citet{Lutsko18}.   The production of equatorial superrotation shifts the phase speed of the equatorial Kelvin waves eastward, potentially damping the interaction with the off-equatorial Rossby waves and damping the instability.  This leads to an intermittency in the eddy momentum flux producing the superrotation, as found by \citet{Mitchell_Vallis10} and explored by \citet{Zurita-Gotor_etal22} and \citet{Nicolas_Vallis26}.

\subsubsection{Tidally-locked planets}
Direct measurements of zonal wind are very difficult to make for exoplanets; however, the phase-curve of radiation received from the combined planet and host star system suggests that, at least in some cases, the hottest region is offset from the substellar point (the point on the equator directly below the host star)  in the direction of the planetary rotation \citep[as described by][and others since]{Knutson_etal07}.   If the offset is due to advection then superrotation may be inferred, and this was explored using a shallow-water model by \citet{Showman_Polvani11} and  then  by \citet{Hammond_Pierrehumbert18}, who showed that standing waves are also important. 

%\begin{figure}
%    \centering
%    \includegraphics[width=\textwidth]{figsuper/Demory}
%    \caption{Demory pattern\dots}
%    \label{fig:Demory}
%\end{figure}    

\begin{figure}
    \centering
    \includegraphics[width=0.8\textwidth]{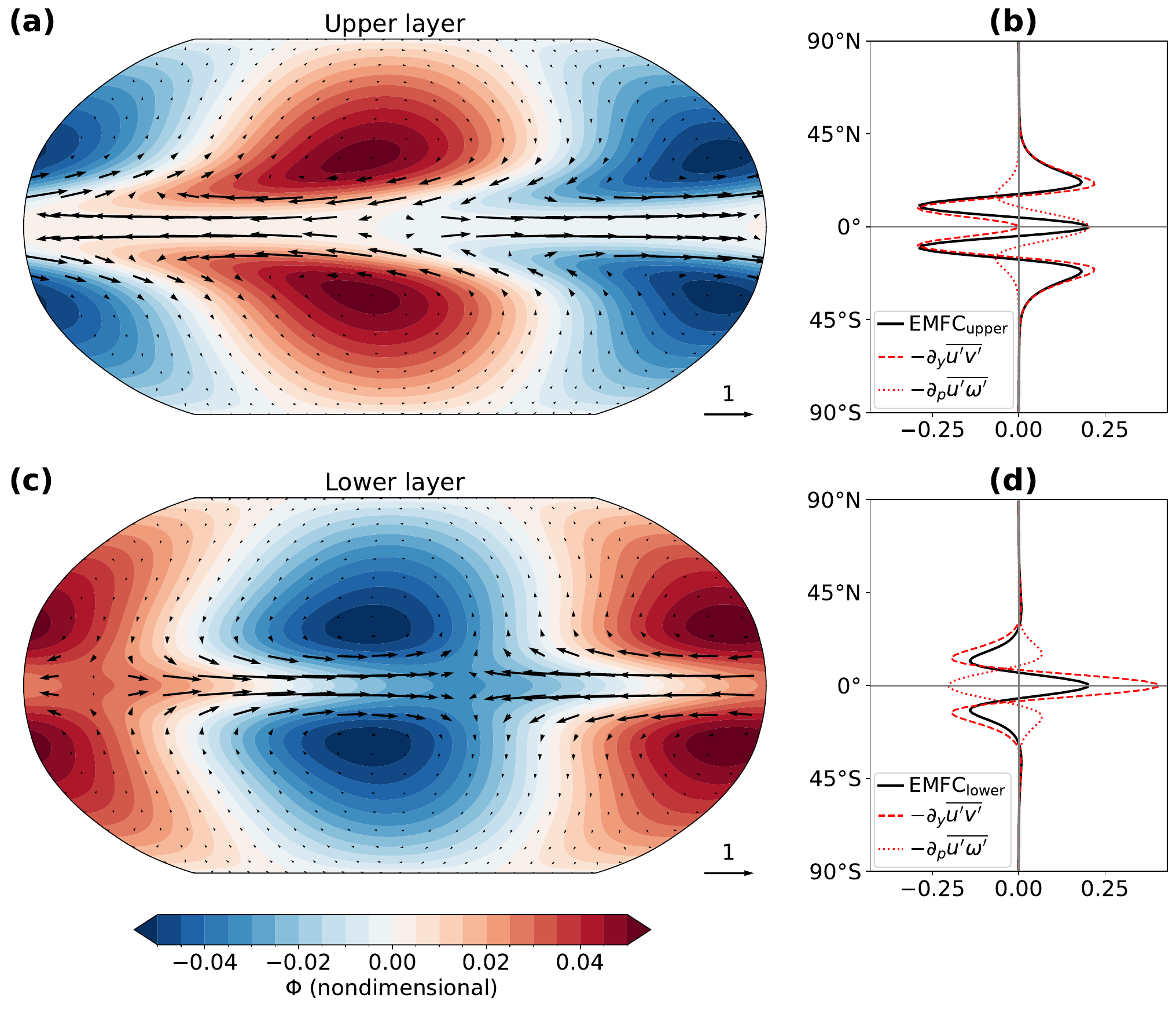}
    \caption{Matsuno–-Gill pattern produced by solar irradiation on a tidally-locked spherical planet using a two-level primitive equation model. The left-hand panels show geopotential and horizontal wind perturbations, as in \figref{fig:RK1}. The eddy flux convergences shown in the right-hand panels indicate momentum flux convergence at the equator, which leads to superrotation in a time-evolving model.}
    \label{fig:MG}
\end{figure}

In a tidally-locked planet the source of heating is stationary and concentrated around the equator, suggesting that the response is likely to be something resembling the Gill pattern \citep[][see  \figref{fig:MG}]{Gill80}, albeit with parameters more appropriate to a slowly-rotating planet.  (\citet{Matsuno66} derived a similar pattern using a periodic forcing, and if no differentiation between the two patterns is needed they are commonly referred to as Matsuno--Gill (MG) patterns.) The basic structure is an equatorial Kelvin mode extending eastward  from the heating and two off-equatorial Rossby waves extending westward.  The pattern  is symmetric across the equator and, therefore, a single-layer model cannot produce superrotation. However, its characteristic chevron shape suggests that with some modification momentum convergence toward the equator should occur,  as with RK instability, and adding vertical structure to the problem does lead to converging eddy momentum fluxes and superrotation (\citealt{Tsai_etal14}; \citealt{Hammond_etal20}).

\begin{figure}
    \centering
    \includegraphics[width=1.05\textwidth]{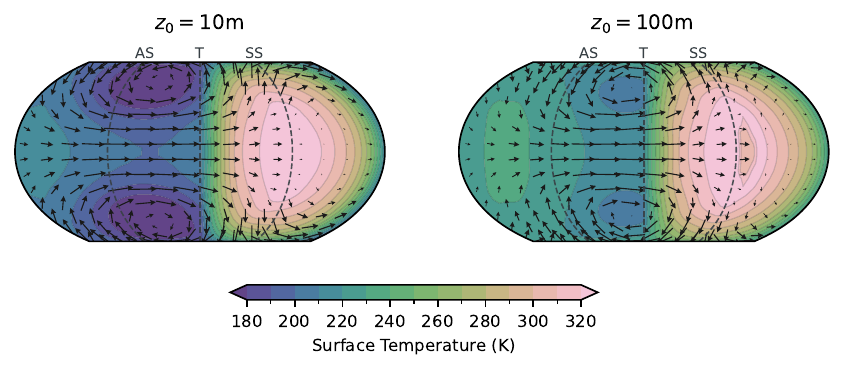}
    \caption{Temperature (colors) and wind (vectors) in two GCM simulations of a moist, Earth-like planet with an active hydrology cycle,  tidally-locked with a rotation period of about 10 days, and inventories of water equivalent to depths of 10 m and 100 m.  The temperatures are at the surface and the velocity vectors are  at 500 hPa.  From left to right, the three faint dashed meridional lines are the anti-stellar, terminator, and substellar longitudes. From \citet{Williams25}.  }
    \label{fig:Williams}
\end{figure}

The MG pattern produced by solar irradiation on a two-level tidally-locked spherical planet is illustrated in \figref{fig:MG}. The lower-layer structure has the canonical chevron-shaped structure, which leads the horizontal flow to converge prograde momentum at the equator. The upper-layer pattern is very different due to the absence of drag, but prograde momentum flux convergence is carried there vertically from the lower layer.  Numerical integrations of the full equations of motion show that, at least in some parameter regimes, superrotation can be produced.  If the Rossby number is reasonably small,  the superrotation can be traced back to the eddy fluxes produced by the \MG model.   If the Rossby number is large a quasi-linear approach is less successful -- an exploration of the parameter regime is given in \citet{Nicolas_Vallis26}.  Superrotation in more complete models  --- for example multi-layer GCMs with a full radiation scheme and hydrology cycle --- also robustly appears, as illustrated in \figref{fig:Williams}, where the hotspot is shifted slightly east of the substellar longitude.

However, superrotation does not always arise, and two factors that do affect its appearance are the  strength of low-level drag and the optical thickness of the atmosphere, as discussed by \citet{Nicolas_Vallis26}.  Drag is important because it affects the phase relation between the lower and upper levels, and hence the crucial vertical transfer of momentum, and without some form of surface drag superrotation is less likely to occur. Similarly, optical thickness is important  because it affects the phase relation between the thermal forcing from the host star and the response.

\subsection{Superrotation and Jets in Deep Atmospheres} \label{sec:deep}

Giant planets, at least those in our Solar System, differ considerably from terrestrial planets in having no clear demarcation between the gaseous atmosphere and a solid planet beneath.  The gas giants, Jupiter and Saturn, also superrotate and have multiple extra-tropical zonal jets whereas the ice giants, Uranus and Neptune, neither superrotate nor have multiple jets.\footnote{The terms `gas giants' and `ice giants' are both misnomers, as the respective planets are neither predominantly gaseous nor icy. Nevertheless, the names are embedded in the community.}
Here we focus on the gas giants, and consider the mechanisms that might give rise to these jets,  and we then briefly speculate as to why the jets on the ice giants are quite different. 

As noted in the Introduction,  there has been something of a debate in the community over the last few decades as to whether the observed jets on Jupiter and Saturn are `deep' or `shallow', with the former meaning that the jets extend into (or originate from) a significant depth in the interior, requiring a full treatment of the Coriolis force with non-hydrostatic equations of motion.    Inversions based on gravimetric observations \citep{Kaspi_etal18} suggest that the jets are aligned with the axis of rotation and extend into the interior about 3000 km in the radial direction, although variations of this depth with latitude are possible (and in particular the extra-tropical jets could be shallower).    Now, flow in shallow atmospheres feels a $\beta$-effect because of the variation of the vertical (i.e., radial, aligned with gravity) component of the rotation vector with latitude. No such effect occurs in the same way in deep flows,  but if the planet is rapidly rotating and in a spherical shell, then the first effect causes the topology of any convection cells to be aligned with the rotation vector (the Taylor--Proudman effect), and the geometry of the shell means that the vertical extent (i.e., in the axial direction) of the convection varies with latitude, giving rise to the topographic $\beta$-effect (\figref{fig:geometry}).
%  giving rise to the presence  of Rossby waves and potentially zonal jets.

A number of authors have explored these effects, perhaps beginning with  \citet{Busse70, Busse76} and with later numerical and theoretical contributions by \citet{Brummell_Hart93,  Yano_etal05,  Busse02}, \citet{Julien_etal06} and others, with \citet{Jones_etal03} finding multiple jets in a Brummel-like model  applied to Jupiter.  A key consideration is the following:  if the strong rotation constrains the motion to such a degree that axial variations of the horizontal flow (meaning flow orthogonal to the rotation axis) are small then some kind of rotating shallow water or quasi-geostrophic approximation may be made.   The $\beta$-effect arising from the variations of the vertical extent of the flow can in principle give rise to Rossby-like waves and these in turn can give rise to zonal jets, and possibly a zonostrophic regime, in a manner similar to that for shallow atmospheres.  However, the precise way in which this works is unsettled, as is the importance or otherwise of the weather layer atop the convective layer, as we now discuss.

\subsubsection{Mechanisms and details}
\begin{figure}
    \centering
    \includegraphics[width=0.8\textwidth]{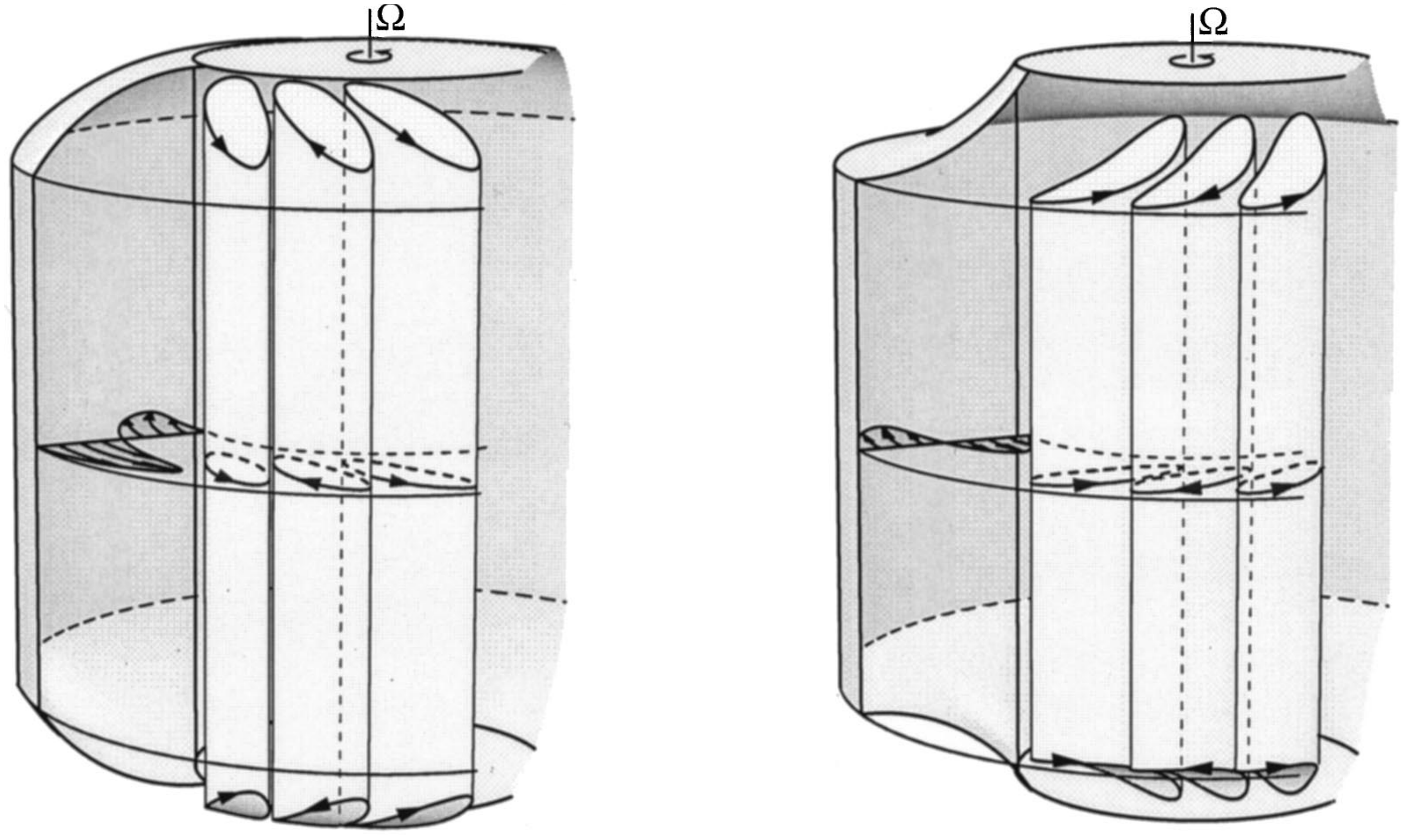}
    \caption{Schematic, from \citet{Busse02}, of  the Influence of curved conical boundaries on the zonal flow. In the convex case the columns tend to spiral outward in the prograde direction thus create a differential rotation with higher angular velocity on the outside
    than on the inside, with the reverse occurring with concave boundaries.}
    \label{fig:Busse}
\end{figure}

Although a sloping boundary is sufficient to produce a topographic $\beta$-effect, the particular way that the boundary slopes may be important for the nature and direction of the jets produced.  Specifically, Busse argues that the convex nature of the boundary of the convecting flow is a key aspect of the mechanism,  for here the  columns tend to spiral outward in the prograde direction, producing a flow with higher angular velocity nearer the outside of the column; i.e., nearer the equator (\figref{fig:Busse}).  The reverse would occur if the boundary were concave.  In this picture the prograde motion at the equator, seen on Jupiter and Saturn,  depends on the tilt of the Taylor-column cross-sections. The mechanism is one of three-dimensional angular momentum transport,  involving components both in spherical radius and latitude. Because the equator is further from the rotation axis, in the convex case  cylindrically outward angular momentum transport is manifested as a prograde flow in equatorial regions.  If for some reason the upper boundary were concave in equatorial regions then subrotating equatorial flow might be expected, as is seen on Uranus and Neptune,  although other mechanisms for subrotation are certainly possible.

Any such jets produced by this or a similar mechanism can be expected to be in good thermal wind balance, given the low Rossby number of  the giant planet atmospheres (at least in the Solar System), although the degree to which the temperature gradient shows any alternating pattern will depend on the vertical structure of the jets and how they decay with depth.  (The Solar System giants have a weaker meridional temperature gradient than Earth, and a larger rotation rate,  so the prograde flow due to  thermal wind balance with the pole--equator temperature gradient alone will be relatively weak compared to the individual jets.)

A Busse-like mechanism is also believed to drive the solar differential rotation, for the Sun exhibits a relatively fast equatorial motion  \citep{Howe09} and could thus be considered `superrotating'. Fully three-dimensional, turbulent simulations of angular momentum transport in the stellar context have commonly reproduced solar-like differential rotation (\citealt{Gilman83}, \citealp{Brun_Toomre02, Matilsky_etal19, Matilsky_etal20},. \citealt{Camisassa_Featherstone22}), and figure 4 of \citet{Gilman83} is a schematic of the mechanisms of superrotation not unlike that of \figref{fig:Busse}.  However, the correspondence between stellar differential rotation and planetary superrotation (with brown dwarfs a useful intermediary) remains to be fully explored.

A seemingly different explanation of the appearance of zonal flows arises from a potential vorticity perspective, if one makes the (perhaps extreme) simplification that the shell of convecting fluid may be treated using the shallow water equations (\citealp{Yano_etal05}, \citealp{Vallis19}).  The   potential vorticity, $Q$, in such a system is given by
\begin{equation}
	\label{PV.1} 
   	Q = {\zeta^* + 2 \Omega \over H},
\end{equation}
where $\zeta^*$ is the axial relative vorticity (i.e., that aligned with the axis of rotation) and $H$ is the depth of the fluid layer in the axial direction, as in \figref{fig:geometry}. The rotation rate, $\Omega$ is constant but $H$ varies with distance from the axis of rotation.
and so with latitude, giving a topographic  $\beta$-effect. To see this, assume that the Rossby number is small and that variations in $H$ occur on a larger scale than variations in $\zeta^*$. The evolution equation for $Q$ then becomes
\begin{equation}
	\label{PV.32} 
   	\DD Q \approx  \frac 1 H \DD {\zeta^*} + 2 \Omega \DD{} \bfrac1H =  \frac 1 H \DD {\zeta^*} - {2 \Omega \over H^2} \vb \cdel H = \frac 1 H \DD {\zeta^*} + \beta^* v,
\end{equation}
where $\beta^* = - (2 \Omega /H) \ppp H y$ and $v$ is the velocity in the $y$-direction (see \figref{fig:geometry}). Evidently,  $\beta^*$ is negative outside the tangent cylinder, positive within it.

\begin{figure}
    \centering
    \includegraphics[width=\textwidth]{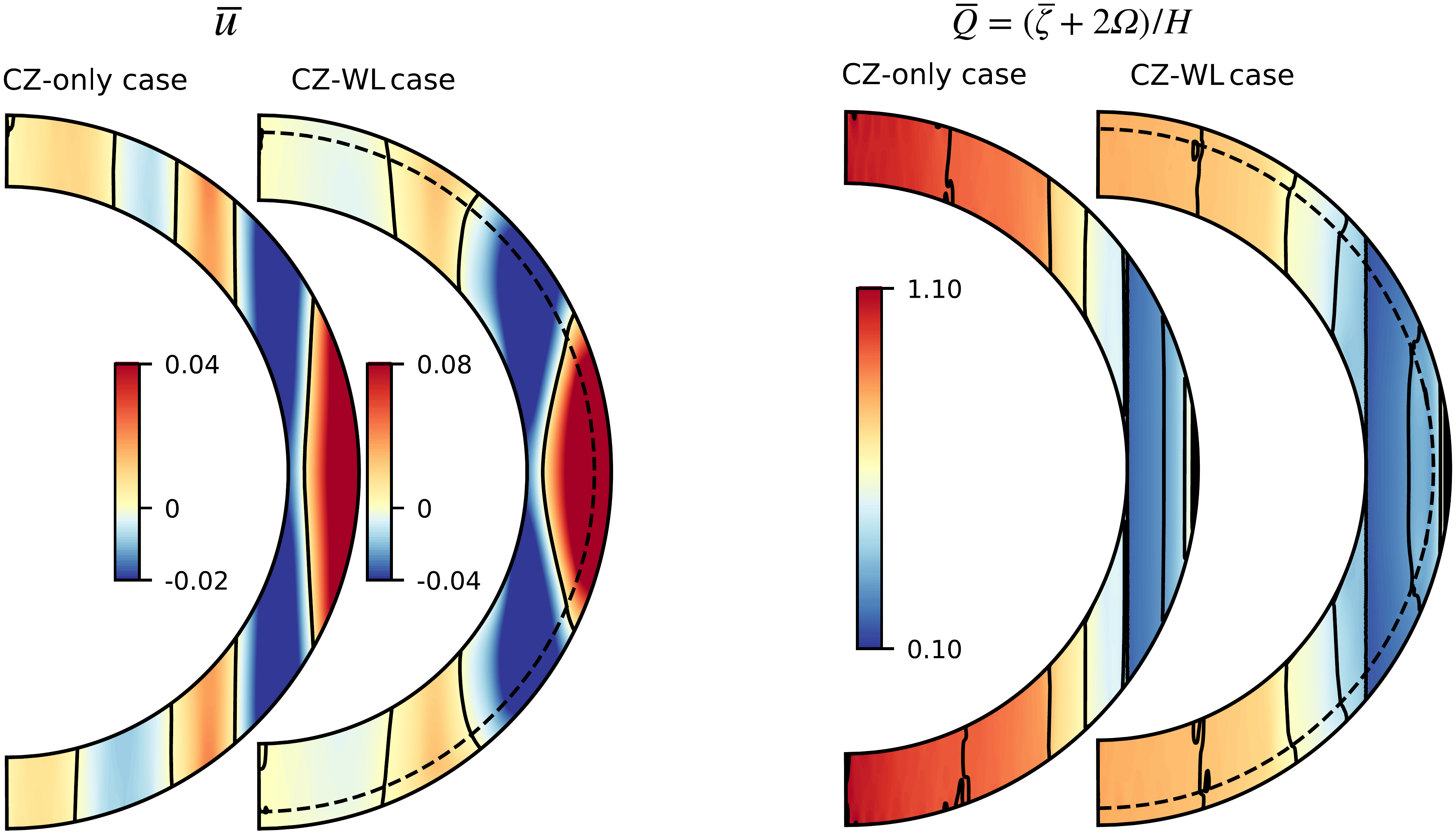}
    \caption{Results from two three-dimensional simulations of the anelastic equations in a spherical, rotating shell forced by heating at the inner boundary and cooling at the top, with Jupiter-like parameters.  `CZ-only' means the simulations have only a convective zone and `CZ-WL' means a stratified, weather-layer zone is imposed at the top, with the dashed semicircle denoting the base of the weather layer. The two left plots show the zonally-averaged zonal velocity, $\obar u$.  The two right plots show potential vorticity, defined as $\overline Q = (\overline{\zeta^*}   + 2\Omega)/H$ (see text and \figref{fig:geometry}). $\overline Q$ is partially homogenized in the region outside of the tangent cylinder, and a stepped structure is apparent inside the tangent cylinder.    Simulations from \citet{Matilsky_etal26}.} 
    \label{fig:pv-jets}
\end{figure}

If the potential vorticity outside the tangent cylinder is homogenized then a straightforward calculation \citep[][Chapter 13]{Vallis19} shows that the zonal flow will be a maximum at the equator.   Three-dimensional numerical simulations of convection in deep spherical shells in a giant-planet setting  do commonly give rise to prograde motion outside the tangent cylinder \citep[e.g.,][]{Heimpel_Aurnou07, Wulff_etal22};  furthermore,  high-resolution simulations run for thousands of rotation periods show partial potential vorticity homogenization accompanying the prograde motion (\figref{fig:pv-jets}).  The homogenization is not complete and occurs on the zonally-averaged quantity $\overline Q$.   The variation of the height $H$ with cylindrical radius, and so with latitude, is crucial here; if $H$ is taken to be constant in the evaluation of  $Q$, homogenization of the potential vorticity does not occur.  In the extra tropics simulations of this nature can produce multiple alternating jets, although there is a tendency for the extra-tropical jets to merge and/or seemingly drift poleward over a very long time period (\citealt{Takehiro_etal24}, \citealt{Matilsky_etal26}).   The mechanism for this is not fully understood, but may be related to the fact that these simulations use a Laplacian viscosity and have insufficient drag to remove energy at large scales, either at the top or bottom of the convective layer, although the origin of  any such drag is not known.  Relatedly,  \citet{Rotvig_Jones06} found multiple jets in simulations with bottom friction but not in simulations with Laplacian viscosity. 

The potential-vorticity homogenization mechanism makes no explicit reference to the convexity or concavity of the bounding surface; rather, prograde equatorial motion arises from the homogenization with a negative $\beta$-effect, and the relation between the mechanisms is not entirely understood.  Fully three-dimensional simulations are expensive and currently prohibit the full exploration of different shapes of bounding surfaces.  Two dimensional simulations do allow one to impose varying types of $\beta$-effect, and if a $\beta$-effect is imposed that mimics the topographic $\beta$-effect of the gas giant planets a prograde, superrotating jet is found in equatorial regions \citep{Zeng_etal25}.  If the region outside of the tangent cylinder is greater (in meridional extent) than the relevant jet scale (e.g., an appropriate Rhines scale) then multiple jets can be expected to form in that region.  This effect can be reproduced in two-dimensional simulations and there is some evidence for this in \figref{fig:pv-jets}.   However, the effect  seems not to be present on the gas giants in the Solar System; that is, the superrotating region seems to occupy the full region outside the tangent cylinder.  That said,  there is some uncertainty as the radial thickness of the convecting shell and the latitudinal extent of the tangent cylinder is thus not precisely known.  Also, Jupiter  does exhibit  distinct peaks of prograde flow within the equatorial region,  perhaps suggesting weaker and possibly shallow alternating jets superimposed on the deeper prograde flow. 

%\begin{SCfigure}
%    \centering
%    \includegraphics[width=0.7\textwidth]{figsuper/Matilsky-fig2}
%    \caption{Zonal flow in 3D simulations of the anelastic equations in a spherical, rotating shell forced by heating at the inner boundary and cooling at the top, with Jupiter-like parameters.  `CZ-only' means the simulations have only a convective zone and `CZ-WL' means a stratified, weather-layer zone is imposed at the top. Superrotation is present in both simulations, with the jets tilting in a radial direction in the weather layer.}
%    \label{fig:MatilskyU}
%\end{SCfigure}    
%
%\begin{SCfigure}
%    \centering
%    \includegraphics[width=0.7\textwidth]{figsuper/Matilsky-fig3}
%    \caption{Potential vorticity in the same simulations as in \figref{fig:MatilskyU}.  Potential vorticity, defined as $(\zeta+2\Omega)/H$ where $\zeta$ is the vorticity in the axial direction and $H$ is the shell height in that direction (as in \figref{fig:geometry}), is well homogenized in the region outside of the tangent cylinder}
%    \label{fig:MatilskyPV}
%\end{SCfigure}    

\subsubsection{Effects of a weather layer}

The visually observed  jets in the Solar System gas giants reside in an upper, stratified, weather layer,  but gravimetric inversions suggest that the jets extend well below that layer.  Thus, some interaction between the convective and stratified layers is expected.  If the solar differential heating in the weather layer gives rise to baroclinic instability, and thence geostrophic turbulence  in a differentially-rotating background state,  then jets can  be excited there independently of the convectively induced ones,  although there are less certain mechanisms for the production of superrotation. (These possibilities are discussed more in \secref{sec:uncertain}).  

In the absence of baroclinic instability in the weather layer, three-dimensional simulations by \citet{Matilsky_etal26} of a stratified layer above a convective layer suggest that one significant effect of the weather layer is to tilt the flow structures so that zonal flow contours become radially aligned rather than axially aligned (\figref{fig:pv-jets}).  The shallowness of the weather layer and the low density of the fluid within it suggest that their gravitational signature of this effect will be weak, and so this result is not inconsistent with observations.   Even if the tilting occurs mainly inside the weather layer it does have an effect deeper into the convective layer,  although the convective cells do remain largely aligned with the rotation axis.  The region in the weather layer is convectively stable (by construction of the model) yet the convective overshoot enables the jets to penetrate well into that layer.  All things considered, the very robust production of convectively forced jets, taken with the results from inversions, does suggest that deep convection has a significant influence in the weather layer and may be the primary cause of the observed jets.  This said, and given that simulations of the weather layer \textit{without} any deep underlying convection can produce zonal jets, it should be said that aspects of the dynamics remain unsettled, as we now discuss. 

\section{Uncertainties, Realities, and Open Questions}
\label{sec:uncertain}

Various mechanisms for jet production and superrotation are now becoming established, but it is less clear which mechanism applies, and under what circumstances, in the different planetary atmospheres.
On Earth-like atmospheres (and in particular on Earth) the extra-tropical eddy-driven jets arise from the action of Rossby-like waves, superimposed on a broader prograde flow arising from thermal wind balance in an atmosphere with a pole--equator temperature gradient.  The processes in Earth's atmosphere are certainly more complicated than \figref{fig:eddyforc} implies, with baroclinic instability, shear of the background flow,  the influence of the subtropical jet and fully nonlinear effects all playing a role.  Furthermore, even with the detailed observations and re-analysis (i.e., state estimates) we have for Earth,  direct unambiguous evidence of Rossby wave activity is hard to come by. Nevertheless,  a $\beta$-effect caused by a potential vorticity gradient in a shallow, stratified atmosphere almost certainly plays a role in the jet-producing mechanism.

In slowly-rotating terrestrial atmospheres,  simulations show that the Rossby-Kelvin instability provides a robust mechanism for superrotation.  We have no clear,  direct observations of the instability in either of the planetary bodies (namely Titan and Venus) in the Solar System for which the mechanism seems likely to apply. Furthermore, Venus is strongly affected by tidal forces and their effect on the vertical transport of momentum is likely to be important.    Numerical simulations of Venus vary widely in their results, making it difficult to determine what the precise mechanism is,  or whether there is a single dominant one. 

In tidally-locked shallow atmospheres a mechanism that  involves the interaction of Rossby and Kelvin modes also seems likely to be important in producing superrotation,  but direct evidence is unobtainable and will be for some time. Indirect evidence of superrotation,  such as the degree of offset of the hotspot relative to the substellar point,  or possibly a measure of the meridional temperature gradient,  may nevertheless give some guidance.   The presence of superrotation also depends on the physical properties of the atmosphere and it seems possible  that, if and when confidence in numerical  simulations is established, the simulations and theory can be used to help characterize the physical properties of the planetary atmosphere.  A number of simulations have highlighted the importance of drag in producing superrotation, and this brings in a major uncertainty: many tidally-locked planets are gas giants (although some are super-Earths) and gas giants have no solid layer that might provide such a drag.  Also,  our theories regarding the response to the thermal forcing are largely based around shallow atmospheres.  This may partially be justified as the solar forcing greatly exceeds the internal thermal forcing (on tidally-locked planets), but the influence of  the planetary interior remains largely unexplored. 

A number of outstanding issues remain regarding the giant planets in the Solar System, with a primary example being the nature of the interaction between the stratified weather layer and the deeper convective layer. It seems very likely that convective motion in the interior does produce zonal flows, and that these manifest themselves in the weather layer, but the degree to which the weather layer can produce jets \textit{independently} is not known with any certainty.  Many simulations of the weather layer alone do  produce jets with a similar character to that of the gas giants (beginning with \citealt{Williams78} and more recently \citealt{Boissinot_etal24}), although the production of superrotation is less common. The kind of superrotation that is seen on Jupiter and Saturn is robustly reproduced in three-dimensional simulations as a consequence of a rapidly rotating convection in the presence of negative $\beta$-effect, and two-dimensional simulations show a similar effect if the convection is parameterized.  Although that effect is very natural,  simulations of Jupiter's atmosphere with shallow-atmosphere or shallow-water models can also give rise to superrotation (e.g., \citealt{Schneider_Liu09}, \citealt{Lian_Showman10}, \citealp{Boissinot_etal24}).  The results depend on the parameters chosen,  in particular those representing mechanical drag, thermal damping and moist effects; however, this does not preclude Saturn and Jupiter being in a parameter regime where superrotation \textit{is} produced by effects in the weather layer, with deeper jets either being a downward extension of the weather layer jets or independent of them.  The zonal flows on the sub--ice-layer  global oceans of the icy moons such as Europa and Enceladus may well resemble that of the gas giants, in that it is convectively forced from below. If so superrotation seems likely since the geometry, with regions inside and outside of a tangent cylinder, is similar, and some simulations reproduce such flow \citep[e.g.][]{Bire_etal22}.  However, complications abound  --- the effects of salinity, the possible internal heating of the water itself, an uneven upper surface --- and observations are very limited.  An  observationally verified understanding of the flows on such bodies seems unlikely to occur in the foreseeable future, but the prospect that they may be incubators of life means interest in them will remain.  

The lack of multiple jets on the ice giants (Uranus and Neptune) and their lack of superrotation suggest that their dynamics are quite different from those of the gas giants and various suggestions have been made as to what the essential difference is  (e.g., \citealt{Liu_Schneider10}, \citealp{Duer_etal25, Zeng_etal25}).  The internal heat flux of Neptune is an order of magnitude smaller than that of the gas giants, and that of Uranus much smaller still,  suggesting that the small scale convective forcing that might stochastically excite multiple jets is weak, yet the zonal flow itself is strong.  If the main forcing of the flow is at a large scale (either internal or from solar forcing)  and the flow is shallow, then subrotating equatorial flow, with broad prograde flows in higher latitudes is more likely than superrotating flow, certainly  in the absence of any mechanism generating  Rossby waves at the equator.   However, the internal structures of Uranus and Neptune are comparatively poorly known \citep{Helled_Fortney20}, and in particular little is certain about the depth of the convective layer (assuming such a layer exists) or whether there is any meaningful concept of there being a different  regions inside and outside of a tangent cylinder.  Evidently, we are at the early stages of understanding how the flow on the ice giants is maintained and we will not discuss them further in this article.

\section{Conclusions}

Let us conclude by summarizing what we can say, with some  confidence, about jets and superrotation in planetary atmospheres.  On rapidly rotating planets (i.e., $\Ro < 1$) with shallow atmospheres, of which we take Earth to be such,  one or more jets in mid-latitudes can be formed by quasi--two-dimensional mechanisms involving turbulence and/or Rossby waves, as adumbrated by \citet{Thompson71}, \citet{Rhines75, Vallis_Maltrud93} and others since.   There are natural mechanisms for the production of the required waves and eddy flow --- baroclinic instability is one --- and on Earth the resulting mid-latitude jet is commonly known as the eddy-driven jet.  This jet is  superimposed on a broad prograde flow that is in thermal wind balance with the meridional temperature gradient.    Equatorial superrotation can also be produced by Rossby waves, and although the mechanisms for equatorial Rossby wave production on giant planets are less well established remains a candidate mechanism for the production of superrotation on Jupiter and Saturn.

Multiple mid-latitude jets can be produced by such mechanisms, and the rapidly-rotating gas giant planets in our Solar System (i.e., Jupiter and Saturn, for which the Rossby number is much smaller than in Earth's atmosphere) do exhibit multiple jets in their midlatitudes, as well as equatorial superrotation.   However, inversions of gravitational data suggest that the jets extend, or originate from, a few thousands of kilometers deep in the interior on Jupiter and even deeper on Saturn, certainly in lower latitudes.  The interior heating in these planets is likely to give rise of convection,  and hence to flow that is three-dimensional in nature. However,  in the presence of rapid rotation in a spherical shell the flow can be approximately described by equations that are quasi-two-dimensional, resembling the familiar quasi-geostrophic equations. The geometry (in particular the sloping boundaries at the top and bottom of the shell) then gives rise to a topographic $\beta$-effect which leads to the production of zonal jets \textit{and} superrotation.  These mechanisms are quite robust and have been reproduced in a number of numerical simulations.  The effect of a stably stratified weather layer lying above the convection means that truly shallow mechanisms cannot be ruled out as having an important effect on the observed jets, especially in mid- and high latitudes.  A reasonable conjecture, consistent with observations and numerical simulations, is that on rapidly rotating gas giants equatorial superrotation is produced by predominantly deep dynamics whereas in mid-latitudes shallow, weather-layer dynamics play an important and possibly essential role. 

In slowly-rotating shallow atmospheres an instability involving the interaction of Rossby and Kelvin waves provides a robust mechanism for the production of prograde flow at the equator, and this is a candidate mechanism for the production of superrotation on Titan and Venus. (This mechanism may not be the whole story,  for example on Venus thermal tides may play a role.)  On tidally-locked planets the interaction of Rossby and Kelvin modes --- now directly forced by stellar insolation --- also provides a mechanism for superrotation.   In numerical simulations the mechanism is found to be robust, although it does not always produce superrotation and not all tidally-locked planets can be expected to superrotate.  Two other caveats are that direct observational verification of the mechanism is obviously hard to come by, and many tidally-locked planets are gas giants with a potentially deep structure.  Little is known about the internal dynamics of slowly-rotating deep planetary atmospheres, tidally-locked or otherwise.

\ack{The authors would like to acknowledge discussions with Matt Browning, Nicholas Brummel, Glenn Flierl, Wanying Kang,  Neil Lewis,  Jonathan Mitchell,  Heng Quan,  Adrian Van Kan, and Yaoxuan Zeng.  For  the purpose of open access the authors have applied a Creative Commons Attribution (CC BY) license to any Author Accepted Manuscript version arising from this submission.}

% \bibliography{super}

\end{document}